\documentstyle[preprint,prb,aps]{revtex}
\newcommand{\bmp}{{\mbox{\boldmath $p$}}}
\newcommand{\bms}{{\mbox{\boldmath $s$}}}
\newcommand{\bmsi}{{\mbox{\boldmath $\sigma$}}}
\newcommand{\bmsna}{{\mbox{\boldmath $\nabla$}}}

\newcommand{\bmF}{{\mbox{\boldmath ${\cal F}$}}}

\newcommand{\bmr}{{\mbox{\boldmath $r$}}}
\newcommand{\bmz}{{\mbox{\boldmath $z$}}}

\newcommand{\bmq}{{\mbox{\boldmath $q$}}}

\newcommand{\Ain}{{\rm {\AA}^{-1}}}

\begin{document}
\preprint {WIS-98/14 June-DPP}
\draft

\date{\today}
\title{Beyond the binary collision approximation for the
large-$q$ response of liquid $^4$He}
\author{A.S. Rinat and M.F. Taragin}
\address{Department of Particle Physics, Weizmann Institute of Science,
         Rehovot 76100, Israel}
\maketitle
\begin{abstract}

We  discuss corrections  to the  linear  response of  a many-body  system
beyond the  binary collision approximation.   We first derive  for smooth
pair interactions  an exact expression  of the response  $\propto 1/q^2$,
considerably simplifying existing forms  and
present  also   the  generalization  for  interactions   with  a  strong,
short-range repulsion.  We then apply the latter to the case of  liquid
$^4$He. We display  the  numerical  influence of  the  $1/q^2$
correction around the  quasi-elastic peak and in  the low-intensity wings
of the  response, far  from that  peak.  Finally  we resolve  an apparent
contradiction in  previous discussions  around the fourth  order cumulant
expansion coefficient. Our results prove that  the large-$q$  response of
liquid $^4$He can be accurately understood on the basis of a dynamical
theory.

\end{abstract} \pacs{}

\section{Introduction}

High-energy pulsed  neutrons from  spallation sources have  recently been
used for the collection of   good-quality   cross sections data for the
inclusive scattering of neutrons from liquid $^4$He. Data are for
temperatures  below  and   above  the  transition  temperature  $T_c$
\cite{sos,and,az}. The above cross sections are a  direct measure of the
dynamic response  or structure  function $S(q,\omega)$,  where $q,\omega$
are the momentum and energy transferred to the system.

It appears convenient  to consider the reduced  response
$\phi(q,y)=(q/M)S(q,\omega)$ instead of $S(q,\omega)$,
where the energy loss parameter $\omega$
is replaced  by an alternative kinematic  variable $y$.  The latter  is a
linear combination of $(q,\omega)$. $M$ is the mass of a
constituent atom.

Virtually all dynamic calculations of the high-$q$ response $S(q,\omega)$
\cite{grs1,sil,ri,rt1,cako,fer,bes,rtmp}  have  been based on  the
Gersch-Rodriguez-Smith (GRS)  expansion of the reduced response in
$1/q\,\,$ \cite{grs1} or  modifications of it \cite{sil,ri,cako,fer,bes}.
The theory in principle employs  only the elementary atom-atom
interaction $V$ and is otherwise free of parameters.

The dominant  part of  the large-$q$ response is  the asymptotic
limit. It describes the response of a neutron striking
an atom with given momentum. The absorption of the
transferred momentum  and energy-loss is  not affected by other  atoms in
the  medium.  Final  State  Interactions (FSI)  induced  by $V$,  produce
corrections to  the above limit  which vanish for increasing  $q$.  The
leading FSI  $\propto   1/q$  is  caused  by  binary
collisions (BC) between a struck and  an arbitrary second particle in the
medium.  For  liquid $^4$He,  predictions for  the reduced
response $\phi(q,y)$ to order $1/q$  agree  well with the data
over a broad  range around the central, quasi-elastic peak  at $y=0$.  In
fact, the quality of the data hardly calls for refinements beyond the BC.
The incentive to  nevertheless consider the introduction  of fine details
is  mainly of  theoretical nature:  A criterion  for the  expansion of  a
function of  two variables $\phi(q,y)$ in  $1/q$ must depend on  $y$.  In
particular  the large  $|y|$ wings  where the  response is  only a  small
fraction of the peak value, has been suspected before to be sensitive
to details beyond the BC.

In  this note  we treat, to  our knowledge  for the  first time, $1/q^2$
corrections.  Those are  due to ternary collisions (TC)  between a struck
and two other particles.  Their study is the major purpose of this note.

A second  topic to be discussed  is related to the  cumulant expansion of
the  response which  has recently resulted in  a successful,
model-independent extraction of the  single-atom momentum distribution in
liquid $^4$He \cite{glyde,glyde1,az1,ken,az2,rtmp}. Our interest here is
the resolution of  an apparent discrepancy between  the directly computed
fourth cumulant coefficient and the value extracted in the BC
approximation for a dynamically calculated response \cite{rtmp}.

We start in Section II recalling some essentials of the GRS expansion for
the  reduced  response,  valid   for  smooth  inter-particle  interaction
\cite{grs1}  and derive  a  formally simple  representation  of TC  terms
$\propto 1/q^2$. Next we mention modifications which are required
if the pair-interaction has a strong, short-range repulsion.
In Section  III we present numerical  values for TC contributions  to the
response of  liquid $^4$He  and discuss its  relative importance,  in the
peak region  and the  low-intensity wings.  Section  IV contains  a brief
discussion  of the  cumulant  representation of  the  response which  has
recently been used to parametrize data for liquid $^4$He.
We  report  a  complete  calculation   of  the  4th  order  FSI  cumulant
coefficient, and thereby also resolve a previously  existing discrepancy.

\section  {Dominant FSI  parts in  the response  for smooth  and singular
interactions.}

Consider the response  $S(q,\omega)$ per particle for  an infinite system
of point-particles

\begin{eqnarray}
S(q,\omega)=A^{-1}(2\pi)^{-1}\int_{-\infty}^{\infty} dt
e^{i\omega t}\langle \Phi_0|\rho_q^{\dagger}(t)\rho_q(0)|\Phi_0\rangle
\label{a1}
\end{eqnarray}

Above $\rho_q(t)$ is the density
operator, translated in time $t$ by the Hamiltonian $H$
\begin{eqnarray}
\rho_q(t)&=&e^{-iHt}\rho_q(0)e^{iHt}
\nonumber\\
\rho_q(0)&=&\sum_j e^{i\bmq.\bmr_j(0)}
\label{a2}
\end{eqnarray}
$\Phi_0$ is the groundstate of the system with energy $E_0$.

We shall work with the reduced response
$\phi(q,y)=(q/M)S(q,\omega)$, where
the  energy loss $\omega$ is replaced by an  alternative
kinematic variable $y=y(q,\omega)$ \cite{grs1,west}
\begin{eqnarray}
y=\frac {M}{q} \bigg(\omega-\frac{q^2}{2M}\bigg )
\label{a3}
\end{eqnarray}
Substitution  of  (2)  into  (1) produces incoherent and coherent
components.   Considering
high-$q$ responses, it suffices to  discuss the dominant incoherent part,
where one tracks a single particle (for instance $'$1$'$) in its
propagation through the medium \cite{west}.

For  the  description of  the  large-$q$  response  we shall  follow  the
formulation of  Gersch, Rodriguez  and Smith  (GRS) \cite{grs1}  and cite
from there a few results.  Details can be found in the bibliography.

It is convenient to introduce the Fourier Transform (FT)  of the  reduced
(incoherent) response
\begin{eqnarray}
\tilde\phi(q,s)=\int^{\infty}_{-\infty} dy e^{-isy} \phi(q,y)
\label{a4}
\end{eqnarray}
The variable $s=(q/M)t$  above is the  distance traveled  in the  medium
during a time $t$ by  a constituent, which moves with constant recoil
velocity $v_q=q/M$: $s$ is the length, canonically conjugate
to the momentum $y$.

The density fluctuations $\rho_q$ in (\ref{a2})  shift
coordinates  in  the   direction  of  $\hat  \bmq$,  chosen   to  be  the
$z$-direction. It leads to the following,  standard expression, valid
for local forces \cite{rahman,grs1} ($\bmr-\bms=\bmr-s\hat\bmq$)

\begin{eqnarray}
\tilde\phi(q,s)&=& (1/A) \bigg
\langle \Phi_0(\bmr_1-\bms;\bmr_k)|
T_{\sigma}{\rm exp}\bigg \lbrace (i/v_q)\int _0^s d\sigma
[H(\bmr_1-\bmsi;\bmr_k)-E_0]
\bigg \rbrace |\Phi_0(\bmr_1;\bmr_k)\bigg \rangle
\nonumber\\
&=&\sum_n(1/v_q)^n\tilde F_n(s)
\label{a5}
\end{eqnarray}
The second line in (5) is the GRS series in $1/v_q$, which is generated
by the expansion  of the above, $\sigma$-ordered exponential.

For local interactions, the Hamiltonian with shifted
coordinate '1 ' can be written as
\begin{eqnarray}
H(\bmr_1-\bmsi;\bmr_k)&=&\sum_l T_l+
\sum_{l\ne 1 \atop k>l } V(\bmr_l;\bmr_k)+
\sum_{k>1}V(\bmr_1-\bmsi;\bmr_k)
\nonumber\\
&=&[\sum_l T_l+
\sum_{{l},{k>l}} V(\bmr_l;\bmr_k)]+
\sum_{k>1}[V(\bmr_1-\bmsi;\bmr_k)-V(\bmr_1;\bmr_k)]
\nonumber\\
&=& H(\bmr_1;\bmr_k)+U_1(\sigma),
\label{a6}
\end{eqnarray}
with
\begin{eqnarray}
U_1(\sigma)(=U_1(\sigma;\bmr_1,\bmr_k))
&=&\sum_{k\ne 1}\delta_{\sigma}V(\bmr_1;\bmr_k)
\nonumber\\
\delta_{\sigma}V(\bmr_1;\bmr_k)
&=&[V(\bmr_1-\bmsi;\bmr_k)-V(\bmr_1;\bmr_k)]
\label{a7}
\end{eqnarray}
$\delta_{\sigma}V(\bmr_1;\bmr_k)$ is the difference between
the interaction of a selected particle $k$ and '1' with the latter,
once at a shifted position $\bmr_1-\bmsi$
and then at $\bmr_1$; $U_1(\sigma)$  is  the  same due to all particles
$k\ne 1$. Using (\ref{a6}) one checks \cite{grs1}
\begin{eqnarray}
[H(\bmr_1-\bmsi;\bmr_k)-E_0]\Phi_0(\bmr_1,\bmr_k)= U_1(\sigma)
\Phi_0(\bmr_1,\bmr_k)
\label{a8}
\end{eqnarray}
It is convenient to  introduce the FT of the GRS coefficient
functions in (\ref{a5}). For example
\begin{eqnarray}
\tilde F_0(s)&=&\frac {1}{A} \int d\bmr_1d\bmr_k
\rho_A(\bmr_1,\bmr_1-\bms;\bmr_k)=
\frac {\rho_1(0;s)}{\rho}
\nonumber\\
&=&\int \frac {d\bmp}{(2\pi)^3} e^{-i\bmp.\bms} n(p)
\label{a9}
\end{eqnarray}
The dominant correction to the above asymptotic limit is
($\bmr=\bmr_1-\bmr_2$)
\begin{eqnarray}
\frac {1}{v_q}\tilde F_1(s)&=& \frac{i}{Av_q} \int d\bmr_1d\bmr_k
\rho_A(\bmr_1,\bmr_1-\bms;\bmr_k) \int_0^s d\sigma [H(\bmr_1-\bmsi)-E_0]
\nonumber\\
&=&-\frac{i}{Av_q}\int d\bmr_1 d\bmr_k\rho_A(\bmr_1,\bmr_1-\bms;\bmr_k)
\int_0^s d\sigma U_1(\sigma)
\nonumber\\
&=&i\int d\bmr
\frac{\rho_2(\bmr,\bmr-\bms;0)}{\rho} \tilde \chi_q(\bmr,s),
\label{a10}
\end{eqnarray}
where use has been made of (\ref{a8}). Above the function $\tilde\chi$
\begin{eqnarray}
\tilde\chi_q(\bmr,s)=-(1/v_q) \int _0^s d\sigma
\delta_{\sigma}V(\bmr),
\label{a11}
\end{eqnarray}
\noindent
is  the  off-shell,  eikonal  phase  in  the  coordinate  representation,
pertinent to the characteristic difference  of interactions acting on '1'
in (\ref{a11}).   In order to  obtain the appropriate on-shell  phase one
needs  to replace  the integration  limits in  (\ref{a11}) from  $0,s$ to
$-\infty,\infty$.

In all,  $\tilde F_1(s)$ describes  binary collisions (BC) of  the struck
particle '1' with  any other particle in the medium.   Its FT $F_1(y)$ is
odd in $y$ and for large $q$ mainly shifts the position of the maximum of
the even asymptotic $F_0(y)$ at $y=0$.

The computation of the above quantities  requires non-diagonal
density matrices. We shall use a normalization, such that
\begin{eqnarray}
\rho_n(1,...n;1',...n')&=&\frac{A!}{(A-n)!}\bigg(\Pi_{j=n+1}^A\,\int d[j]
\bigg )\Phi_0(1,...n;n+1,..n_A)\Phi_0(1',....n';n+1,...n_A)
\nonumber\\
&=&[(A-n-1)]^{-1}\int d[n+1]\rho_{n+1}(1,...n,n+1;1',...n',n+1)
\nonumber\\
\rho_A(1,...A;1',...A')&=&A!\Phi_0(1,...A)\Phi_0(1',...A'),
\label{a12}
\end{eqnarray}
The  densities  $\rho_n(\bmr_1,\bmr_1-\bms;\bmr_k)$,   required  in  Eqs.
(\ref{a9}), (\ref{a10})  , are diagonal  in all particles $k$,  except in
'1'.  For example,  $\rho_1(0;s)=\rho_1(\bmr,\bmr-\bms)$ in (\ref{a9}) is
the single-particle density matrix which has the single-particle momentum
distribution  $n(p)$   as  its  FT. Its  diagonal  part
$\rho_1(0,0)=\rho$ is the number density.

The first two  terms in the GRS series  $F_0,F_1$ satisfactorily describe
the  data  for   the  reduced  response  in  a  broad   band  around  the
quasi-elastic peak at $y=0$.  Such a fit cannot be expected in the wings,
where $ F_1(y)$ occasionally reaches  small negative values.  For growing
$ y $    in    those    wings    $F_l(y),\,l\ge    2$    competes    with
$F_0(y)+(1/v_q)F_1(y)$ of comparable size: ultimately $\phi(q,y)\ge 0$.

Partial  expressions   for  the   next-to-leading  order   terms  $\tilde
F_2(s),F_2(y)$ have been given before by  Gersch et al \cite{grs1} and by
Besprosvany  \cite{bes} but  those forms  are  not complete  and are  not
always  transparent.  We  shall derive  below expressions  for the  exact
$\tilde F_2$, based on Eqs. (12), (15) in Ref. \onlinecite{grs1}
\begin{eqnarray}
\tilde F_2(s)= \frac{i^2}{A} \int d\bmr_1 d\bmr_k
\rho_A(\bmr_1,\bmr_1-\bms;\bmr_k)
\int_0^s d\sigma [H(\bmr_1-\bmsi,\bmr_k)-E_0]
\int_0^{\sigma'} d\sigma' [H(\bmr_1-\bmsi',\bmr_k)-E_0]
\label{a13}
\end{eqnarray}
Consider the operators in the brackets above,
acting on the wave functions which compose $\rho_A$,
Eq. (\ref{a12}), with shifted, respectively unshifted coordinate
$\bmr_1$. Their combined result is
\begin{eqnarray}
\tilde F_2(s)&=& \frac{i^2}{A} \int d\bmr_1 d\bmr_k
\rho_A(\bmr_1,\bmr_1-\bms;\bmr_k)
\int_0^s d\sigma [U_1(\sigma)-U_1(s)]
\int_0^{\sigma} d\sigma' U_1(\sigma')
\nonumber\\
&=&\frac{i^2}{A} \int d\bmr_1 d\bmr_k \rho_A(\bmr_1,\bmr_1-\bms;\bmr_k)
\bigg \lbrace {\frac{1}{2}\bigg \lbrack \int_0^s d\sigma U_1(\sigma)
\bigg\rbrack} ^2 - U_1(s)\int_0^s\,\,d\sigma\,\int_0^{\sigma}
d\sigma' U_1(\sigma') \bigg \rbrace
\label{a14}
\end{eqnarray}
Since $\tilde F_2\propto U^2$ and the latter $\propto V^2$
we dub $\tilde F_2$ the TC contribution to FSI \cite{foot}.

We start with the first term in the braces in Eq. (\ref{a14}).
Using the definition in (\ref{a7}) one rewrites
\begin{eqnarray}
{\bigg \lbrack\int_0^s\,d\sigma U_1(\sigma)\bigg \rbrack}^2
&=&\sum_{k\ne 1}{\bigg \lbrack \int_0^s  d\sigma
\delta_{\sigma}V(\bmr_1-\bmr_k)\bigg \rbrack}^2
\nonumber\\
&& +\sum_{1\ne l\ne k \ne 1}\bigg\lbrack \int_0^s\,d\sigma\,
\delta_{\sigma}V(\bmr_1-\bmr_k)\bigg \rbrack
\bigg \lbrack\int_0^s\, d\sigma\,
\delta_{\sigma}V(\bmr_1-\bmr_l)\bigg \rbrack
\label{a15}
\end{eqnarray}

The above components are distinct two- and three-particle
operators and the same holds for the parallel decomposition
of the second term in the braces in (\ref{a14}). With
$\bar\bmr_1=\bmr_1-\bmr_2;\,\,\bar\bmr_3=\bmr_3-\bmr_2$)
one may reduce (\ref{a14}) to
\begin{mathletters}
\label{a16}
\begin{eqnarray}
\tilde F_2(s)&=&\tilde F_2^{(2)}(s)+\tilde F_2^{(3)}(s)
\nonumber\\
\frac{1}{v_q^2}\tilde F_2^{(2)}(s)&=&\int d\bmr \frac
{\rho_2(\bmr-\bms,0;\bmr,0)}{\rho} \bigg \lbrack \frac {1}{2}
[i\tilde\chi_q(\bmr,s)]^2
+\frac{i}{v_q}\delta_sV(\bmr) \int_0^s
d\sigma [i\tilde\chi_q(\bmr,\sigma)]\bigg \rbrack
\label{a16a}\\
\frac {1}{v_q^2}\tilde F_2^{(3)}(s)
&=&\int d\bar\bmr_1 d\bar\bmr_3 \frac
{\rho_3(\bar\bmr_1-\bms,0,\bar\bmr_3;\bar\bmr_1,0,\bar\bmr_3)} {\rho}
\bigg \lbrack \frac {1}{2}
[i\tilde \chi_q(\bar\bmr_1,s)] [i\tilde \chi_q(\bar\bmr_1-\bar\bmr_3,s)]
\nonumber\\
&&+\frac{i}{v_q}\delta_sV(\bar\bmr_1)
\int_0^s\,d\sigma [i\tilde\chi(\bar\bmr_1-\bar\bmr_3,\sigma)]
\bigg \rbrack,
\label{a16b}
\end{eqnarray}
\end{mathletters}
with $\tilde \chi_q(\bmr,s)$ the off-shell phase as defined in
(\ref{a11}) (Note that $\tilde\chi \propto 1/v_q$).

For smooth,  non-singular local forces, the above completes  the
derivation of an exact expression for $\tilde F_2(s)$. However, if $V$
possesses a strong, short-range repulsion, as is the case for atom-atom
interactions, difficulties emerge. There are no problems if
in integrands wave functions or density matrices and $V$ have identical
arguments, in which case large values of $V$  are generally
off-set by small values of $\rho_n$ at common small $\bmr$.
This is not the case in general. A prime  example is  Eq. (\ref{a10})
with $\bmr $-dependence  through $V(\bmr-\bmsi)$, $0\le |\bmsi|\le s$,
and $\bmr-\bmsi$ generally not coinciding with either $\bmr-\bms$ or
$\bmr$ in $\rho_2(\bmr-\bms,\bmr;0)$: Large line integrals may result.

In the above case smooth expressions emerge again
upon  summation of a ladder of pair interactions
$V(r)$, leading to $V_{eff}$ which is the eikonal, off-shell
$t$ matrix \cite{ri,rt1,bes}. Effectively
\begin{eqnarray}
i\tilde \chi\to {\rm exp}[i\tilde \chi]-1=
i\tilde \chi+1/2 [i\tilde \chi]^2+...
\label{a17}
\end{eqnarray}
Using (\ref{a10}) and (\ref{a17}) we define
\begin{eqnarray}
\tilde G_2(s,[V])&\equiv& \tilde F_2(s,V\to [t])
=\tilde G_2^{(2)}(s)+\tilde G_2^{(3)}(s),
\label{a18}
\end{eqnarray}
with the following two- and three-particle components
\begin{mathletters}
\label{a19}
\begin{eqnarray}
\frac{1}{v_q^2}\tilde G_2^{(2)}(s)&=& \int d\bmr \frac
{\rho_2(\bmr-\bms,0;\bmr,0)}{\rho} \bigg \lbrack
\frac{i}{v_q}\delta_sV(\bmr)  \int_0^s d\sigma \bigg (
{\rm exp}{[i\tilde\chi_q(\bmr,\sigma)]}-1\bigg )\bigg \rbrack
\label{a19a}\\
\frac {1}{v_q^2}\tilde G_2^{(3)}(s)
&=&\int d\bar\bmr_1 d\bar\bmr_3 \frac
{\rho_3(\bar\bmr_1-\bms,0,\bar\bmr_3;\bar\bmr_1,0,\bar\bmr_3)}{\rho}
\bigg \lbrack \frac {1}{2}
\bigg ( {\rm exp}{[i\tilde\chi_q(\bar\bmr_1,s)]}-1 \bigg )
\nonumber\\
&&*\bigg ( {\rm exp}{[i\tilde\chi_q(\bar\bmr_1-\bar\bmr_3,s)]}-1 \bigg )
+\frac{i}{v_q}\delta_sV(\bar\bmr_1) \int_0^s d\sigma
\bigg ({\rm exp}{[i\tilde\chi_q(\bar\bmr_1-\bar\bmr_3,\sigma)]}-1\bigg )
\bigg \rbrack
\label{a19b}
\end{eqnarray}
\end{mathletters}
Care should be exercised in the replacement $V\to V_{eff}$ for an
ill-behaved $V$ as we shall now illustrate by focusing on
the first term in the brackets in
(\ref{a16a}), $\frac{1}{2}[i\tilde\chi]^2$.
It can be shown that $all$ higher order
terms $\tilde F_n(s,[V])$ contain a 2-body component of the form
\begin{eqnarray}
\frac{1}{v_q^n}\tilde F_{n}^{(1)}(s,[V])=\int d\bmr \frac
{\rho_2(\bmr-\bms,0;\bmr,0)}{\rho} \bigg \lbrack
(i\tilde\chi)^n/n!\bigg \rbrack
\label{a20}
\end{eqnarray}
Using  (\ref{a17}) those may be summed up to
\begin{eqnarray}
\sum_{n>1}\tilde F_n^{(1)}(s,[V])=\tilde F_1(s,[t])
\label{a21}
\end{eqnarray}
If therefore $\tilde F_1(s,[V])$  has been
regularized by $V\to V_{eff}=t$, the first part of
$\tilde F_2^{(2)}(s)$ in Eq. (\ref{a16a})
is already contained in $\tilde F_1(s,[t])$ and
in order to avoid double-counting it should be removed
from $\tilde G_2(s,[V])$. We note that by
construction, the remaining term in  Eq. (\ref{a19a}) which emerges
from TC is nevertheless of 2-body character.

Eqs. (\ref{a19}) above are the exact TC
contributions to the GRS series (\ref{a5}) for the reduced response.
At this point we mention an alternative to the
GRS series, namely the cumulant representation for
the FT of the reduced response \cite{grs2}
\begin{mathletters}
\label{a22}
\begin{eqnarray}
\tilde \phi(q,s)&=&
\tilde F_0(s)\tilde R(q,s)=\tilde F_0(s){\rm exp}[\tilde\Omega(q,s)]
\label{a22a}\\
\tilde R(q,s)&=&\sum_{n\ge 1}\bigg(\frac{1}{v_q}\bigg )^n
\frac {\tilde F_n(s)}{\tilde F_0(s)}
\label{a22b}\\
\tilde\Omega &=&(\tilde R-1)-(1/2)(\tilde R-1)^2+...,
\label{a22c}
\end{eqnarray}
\end{mathletters}
with all FSI effects contained in either $\tilde R$ or $\tilde\Omega$

Below we report a calculation of TC contributions,
using choices for the underlying densities.

\section{TC contributions to the response of liquid $^4$He.}

Until now, dynamical calculations based on the GRS series were
limited to FSI in the BC approximation,  i.e.
\begin{eqnarray}
\tilde\Omega\to \tilde\Omega^{BC}=\tilde\Omega_2=
\frac {\tilde F_1}{v_q \tilde F_0}-1
\label{a23}
\end{eqnarray}
with  the  GRS series, cut at  $n=2$,  as in Ref.
\onlinecite{rtmp}. We now report what apparently are the first results
for  the next-to-leading  order TC  corrections and
which  are contained  in
$\tilde G_2=\tilde G_2^{(2)}+G_2^{(3)}$, Eqs. (\ref{a18}), (\ref{a19}).

We first recall  the standard input described  in Ref. \onlinecite{rtmp}.
For  the   bare  interaction  we   use  the  standard   $V^{Aziz}$,  Ref.
\onlinecite{aziz},  and  for the single-atom  momentum  distribution  the
results   of  Refs.   \onlinecite{wp,cep1}.   As   regards  semi-diagonal
2-particle  density  matrix   $\rho_2(\bmr-\bms,0;\bmr,0)$,  there  exist
results obtained using stochastic methods  \cite{cep2,fer}, but
computationally it is unnecessarily time-consuming to evaluate
those for each and every $(\bmr,s)$, as  required  in calculations
of the expressions (\ref{a10}) or (\ref{a19a}).

In the past relatively simple guesses have been made for $\rho_2$.
We shall use below the interpolation formula by GRS \cite{grs1,ri}
\begin{eqnarray}
\rho_2(\bmr-\bms,0;\bmr,0)(=\rho_2(\bmr-\bms,\bmr;0))
&\equiv&\rho\rho_1(0,s)\zeta_2(\bmr-\bms,\bmr)
\nonumber\\
\zeta_2(\bmr-\bms,\bmr)
&\approx& \sqrt{g(|\bmr-\bms|)g(r)}
\label{a24})
\end{eqnarray}
with $g(r)$ the pair-distribution function, chosen to be the one
from Ref. \onlinecite{fer}.

A calculation of $\tilde G_3$ requires the 3-particle density matrix
$\rho_3$ which, as before is non-diagonal in coordinate 1.
As an approximation we suggest
\begin{eqnarray}
\rho_3(\bar\bmr_1-\bms,0,\bar\bmr_3;\bar\bmr_1,0,\bar\bmr_3)
&\approx& \frac{(A-2)}{(A-1)}
\frac {\rho_2(\bar\bmr_1-\bms,0;\bar\bmr_1,0)
\rho_2(\bar\bmr_3-\bms,0;\bar\bmr_1-\bar\bmr_3,0)}{\rho_1(0;s)}
\nonumber\\
&=&\frac{(A-2)}{(A-1)} \rho^2\rho_1(0;s)
\zeta_2(\bar\bmr_1-\bms,\bar\bmr_1)
\zeta_2(\bar\bmr_1-\bar\bmr_3-\bms,\bar\bmr_1-\bar\bmr_3)
\label{a25}
\end{eqnarray}
where use has been made of (\ref{a24}). The choice (\ref{a25}) has
several advantages

i)  With $'1'$  paying a special role, it is symmetric in the other
coordinates

ii) It exactly satisfies the $'$sumrule$'$ (\ref{a12})

iii) It factorizes in parts dependent on
$\bar\bmr_1, \bar\bmr_1-\bar\bmr_3$

iv) It causes $\rho_3$ to vanish for small values of the 4 coordinates
which would otherwise produce large values for the
factors in the operator in the brackets in (\ref{a19b}).

An immediate consequence of iii) above is the reduction of the,
effectively 5-dimensional integral in (\ref{a19b}) to the product of
two,  2-dimensional integrals
\begin{mathletters}
\label{a26}
\begin{eqnarray}
\frac{1}{v_q^2}\frac{\tilde G_2^{(2)}(s)}{\tilde F_0(s)}
&=& \rho \int d\bmr \zeta_2(\bmr-\bms,\bmr) \bigg \lbrack
\frac{i}{v_q}\delta_sV(\bmr)  \int_0^s d\sigma \bigg (
{\rm exp}{[i\tilde\chi_q(\bmr,\sigma)]}-1\bigg )\bigg \rbrack
\label{a26a}\\
\frac {1}{v_q^2}\frac{\tilde G_2^{(3)}(s)}{\tilde F_0(s)}
&=& \rho^2 \bigg (\frac{1}{2}\bigg\lbrack \int d\bar\bmr_1
\zeta_2(\bar\bmr_1-\bms,\bar\bmr_1)
\bigg ( {\rm exp}{[i\tilde\chi_q(\bar\bmr_1,s)]}-1 \bigg )\bigg\rbrack ^2
+\bigg\lbrack \frac{i}{v_q} \int d\bar\bmr_1
\zeta_2(\bar\bmr_1-\bms,\bar\bmr_1)
\delta_sV(\bar\bmr_1)\bigg \rbrack
\nonumber\\
&*&\bigg \lbrack \int d\bar\bmr_3'
\zeta_2(\bar\bmr_3'-\bms,\bar\bmr_3')\int_0^s d\sigma
\bigg ({\rm exp}{[i\tilde\chi_q(\bar\bmr_3',\sigma)]}-1\bigg )\bigg
\rbrack \bigg )
\label{a26b}
\end{eqnarray}
\end{mathletters}

Anticipating small TC corrections we approximate the cumulant
representation (\ref{a22})
\begin{eqnarray}
\tilde \Omega^{TC} &\approx& \tilde R^{TC}-1
\nonumber\\
\tilde R^{TC}&=&\frac {\tilde G_2}{v_q^2 \tilde F_0}
\label{a27}
\end{eqnarray}
The thus defined TC  contribution to the FSI phase has  been added to the
previously calculated  BC part  $\tilde\Omega^{BC}$, Eq.  (\ref{a23}).
From (\ref{a22a}) and the inverse of (\ref{a4}), we compute the response
for $T=2.5$ K to the corresponding order.

A  first  observation  is  the  relative insignificance  of  3-body  TC
contributions for the $q$-range  investigated.  A heuristic argument runs
as follows.  If the BC FSI contributions amounts to a fraction of the the
dominant  asymptotic limit,  one estimates
from  the factorization  (\ref{a25}) of  the
3-body density matrix that TC FSI is approximately the square
of that fraction of $F_0$.

We now display some results for TC contributions.
Figs. 1a,b show for small $y$ and $q=21,25,29,50,100\,\Ain$ the even
part of the calculated reduced
response $\phi^{even}(q,y)=[\phi(q,y)+\phi(q,-y)]/2$ ,without and
including TC contributions (note that  $\phi^{TC}$ is even).  Even for
$y=0$ there  is an effect which for increasing $q\ge 21 \Ain$
decreases from $2{\rm  to}0 \%$.

Figs. 2a,b show the  fractional effect of TC contributions
\begin{eqnarray}
\alpha(q,y)=1+\frac {\phi^{TC}(q,y)}{\phi^{BC}(q,y)}
\label{a28}
\end{eqnarray}
calculated for $2.5 \lesssim |y| \,({\rm in \,\Ain}) \lesssim 3.3$. The
difference in sign of $\alpha-1$ clearly shows the effect of the
competition between the dominant even and odd  parts in the wings
of  the  response.

Finally, Figs. 3a,b,c show for $q=21,25,29\Ain\,,T=2.5$K   the efffect
of TC on the calulated response, including the unresolved
effect of  the instrumental resolution.  The  small TC
contributions discernibly improve the  agreement of predictions with data
in the above $y$ regions.

\section{On the fourth central moment of the response.}

The preceding Sections deal with
the reduced response (\ref{a22}) up to, and including TC
contributions and its, in principle exact, calculation. For the FT
of those one needs $\tilde F_n(s), n \le 2$ or (cf. {\ref{a22}))
$\tilde\Omega(q,s)$ to that order, both for all relevant $s$.

We now address a second topic which is related to the cumulant
representations (\ref{a22}) and which is based on the small-$s$
expansions
\begin{mathletters}
\label{a29}
\begin{eqnarray}
\tilde F_0(s)&=&\sum_{m \ge 2}\frac {(-is)^m}{m!}\bar\alpha_m\
\label{a29a}\\
\tilde\Omega(q,s)&=&\sum_{m \ge 3}\frac {(-is)^m}{m!}\bar\beta_m(q)
\label{a29b}
\end{eqnarray}
\end{mathletters}
The above coefficients $\alpha_m$ are related to even moments of
the momentum distribution $n(p)$, while the FSI coefficient functions
$\bar\beta_m(q)$ in the  expansion (\ref{a29b})
can be  expressed in terms of central moments of the response (see for
instance Ref. \onlinecite{sears}).
\begin{eqnarray}
{\cal M}_n(q)&=&\int d\omega(\omega-q^2/2M)^n S(q,\omega)
\nonumber\\
&=& (v_q)^n\int dy\, y^n\phi(q,y)\equiv (v_q)^n\bar {\cal M}_n(q)
\label{a30}
\end{eqnarray}
For our purpose it suffices to give the following expressions for
$n$=3,4 and valid for local interactions $V$ \cite{plac}
\begin{mathletters}
\label{a31}
\begin{eqnarray}
\bar\beta_3 &=& \bar {\cal M}_3 =
\bigg (\frac{1}{6v_q}\bigg )\langle\nabla^2\,  V\rangle
\label{a31a}\\
\bar\beta_4 &=& \bar {\cal M}_4-\bar\alpha_4-3\bar\alpha_2^2
= \bigg (\frac{1}{3v_q^2}\bigg )\langle \bmF_1.\bmF_1\rangle
\label{a31b}
\end{eqnarray}
\end{mathletters}
\noindent
$\bmF_1\,$ above
\begin{eqnarray}
\bmF_1=\sum_{k\ne 1}\bmF_1(1,k)=-\bmsna_1\sum_{k>1}V(\bmr_1-\bmr_k),
\label{a32}
\end{eqnarray}
distinct from $U_1(s)$, Eq. (\ref{a7}),
is the true total force on a given particle $'$1$'$  \cite{foot1}.
The expectation value in (\ref{a31b}) can thus be separated in two parts.
The first contains the    square of    the
force on $'1'$ due to one particle and in the second part forces on
$'1'$ by two different particles
\begin{eqnarray}
\langle \bmF_1.\bmF_1\rangle=
\langle \sum_{j\ne 1}[\bmF_1(1,j)]^2\rangle+ \langle
\sum_{1 \ne j \ne k\ne 1}\bmF_1(1,j)\bmF_1(1,k)\rangle
\label{a33}
\end{eqnarray}
The expansions (\ref{a29}) provide a parametrization of the response, but
the technique has been shown to  have its problems \cite{rtmp}.  One such
problem is the convergence for growing $s$ which is indispensable for the
calculation of the inverse FT (\ref{a4}) from $\phi(q,y)$.  Moreover, the
cumulant expansion lacks  a systematic ordering in powers  of $1/q$ which
is also remedied in GRS theory.  Notwithstanding, there has recently been
a  renewed interest  in  the  above small  $s$-cumulant  expansions as  a
vehicle  to extract  the  single-atom momentum  distribution $n(p)$  from
response  data for  $^4$He  and Ne  \cite{glyde,glyde1,az1,ken,az2,rtmp}.
Around $\bar\beta_4$  an apparent contradiction arises,  which we discuss
below.

One may $'$invert$'$ Eqs. (\ref{a29}) in order to find
alternative expressions for
$$\bar\beta_m(q)=m!\,i^m\,\lim_{s \to 0}[\tilde\Omega(q,s)/s^m]  $$
In particular
\begin{mathletters}
\label{a34}
\begin{eqnarray}
\bar\beta_3(q)&=&\,\,6\,{\lim_{s\to 0}}\,[{\rm Im} \tilde\Omega(q,s)/s^3]
\label{a34a}\\
\bar\beta_4(q)&=&24\,{\lim_{s \to 0}}\,[{\rm Re} \tilde \Omega(q,s)/s^4]
\label{a34b}
\end{eqnarray}
\end{mathletters}
The above  FSI coefficient functions can  only be calculated if  a theory
provides the FSI  phase function $\tilde\Omega(q,s)$.  The  GRS theory is
one  such example.   The dynamic  calculation described  in the  previous
Sections, provides $\tilde\Omega(q,s)$ for all $s$.

First we state  that without truncations, the cumulant  expansion and the
GRS series ought to  lead to the same response, and  in particular to the
same   numerical   values   for  the   cumulant   coefficient   functions
$\bar\beta_m$.   This  is not  self-evident  since  Eqs. (\ref{a31})  and
(\ref{a34}) look quite dissimilar.  The  former are expectation values in
terms  of   $diagonal$  density  matrices,  whereas   the  $\tilde\Omega$
underlying   the   GRS  theory   is   an   operator,  averaged   over   a
$non$-$diagonal$  density  matrix.   Nevertheless  the  identity  of  the
derived $\bar\beta_3(q)$ has been formally  verified in the past (see for
instance Ref. \onlinecite{fer}).   In addition a numerical  test has been
performed  using  $\tilde\Omega  \to \tilde\Omega^{BC}$,  which  suffices
since   $\tilde\beta_3$   draws   entirely   on  $\tilde   F_1$   or   on
$\tilde\Omega^{BC}$.  The calculated value and  the one, extracted over a
wide $q$ range, indeed agree to high accuracy \cite{rtmp}.

For a similar demonstration regarding
$\bar\beta_4$ one uses (\ref{a22b}), (\ref{a14}), (\ref{a15}) in
(\ref{a34a}) and readily
verifies  that terms  $\propto s^4$,  needed in  the threshold  behaviour
(\ref{a34b}),  originate exclusively  from the  TC terms  $\tilde G_2(s)$
(cf.  Eqs. (\ref{a19})).  Observing that  $\delta V$ in (13), (14) always
appears quadratically, one has
\begin{eqnarray}
\delta_sV(\bmr)&=&(1/2)s^2\frac{\partial V(r)}{\partial z}+{\cal O}(s^3)
\nonumber\\
UU &\propto & s^4(\hat\bmz.\bmF_1^2+{\cal O}(s^5),
\label{a35}
\end{eqnarray}
and (\ref{a34}) results.

We separately treat BC and TC contributions to $\bar\beta_4(q)$ and
start with the above mentioned BC approximation for the regularized
$\tilde F_1(y,[t])$. One observes that even the BC FSI phase
function $\tilde\Omega^{BC}(q,s)$
contains terms $\propto s^4$, contributing to $\bar\beta_4$.
Again, a remarkably stable, $negative$ value could be extracted from
calculated BC phases over a  wide $q$-range \cite{rtmp}
(${q^*}=q/10$ in $\Ain$)
\begin{eqnarray}
{q^*}^2\bar\beta_4^{(21)}(q)=(-2.27   \pm   0.02)  \AA^{-4}
\label{a36}
\end{eqnarray}
One easily demonstrates that the same, $computed$ from  the first term
in the brackets of (\ref{a16a}) is
$q^{*2}\bar\beta_4^{(21})(q)= -(M/10)^2\sum_{j\ne 1}\langle
\bmF_1(1,j)^2\rangle =-2.19  \AA^{-4}$, again in close agreement with the
$extracted$ result (\ref{a36}). The negative outcome  clearly
contradicts the manifestly positive expression
(\ref{a31b}) for the complete $\bar\beta_4$. The latter, however, draws
also on additional TC contributions from $\tilde G_2^{(2)}$ and
$\tilde G_2^{(3)}$, Eqs. (\ref{a19}), which we now address.

We start with
the threshold value of the two-body part $\tilde G_2^{(2)}$
of the TC contribution, which is readily shown to be
exactly 4/3 times the first part and for the positive, complete  two-body
part one finds (cf. (\ref{a33}))
\begin{eqnarray}
q^{*2}\bar\beta_4^{(2)}(q)=
q^{*2}[\bar\beta_4^{(21)}(q)+\bar\beta_4^{(22)}(q)]=
\frac{1}{3}(M/10)^2 \sum_{j\ne 1}
\langle \bmF_1(1,j)^2\rangle=0.73 \AA^{-4}
\label{a37}
\end{eqnarray}
Within $\approx 0.5\%$
the same value  results when calculating the  threshold value (\ref{a36})
and using for the FSI phase function $\tilde \Omega(q,s)$ Eq. (\ref{a22})
with $\tilde  G_2^{(2)}$ as in (\ref{a18}),  (\ref{a19a}).  As emphasized
before,  the  close  agreement   evidences  numerical  accuracy  and  not
consistency.

The genuine 3-body TC part, defined by (\ref{a33}), {\ref{a19b})
\begin{mathletters}
\label{a38}
\begin{eqnarray}
\bar\beta_4^{(3)}(q)&=&
24\,{\lim_{s \to 0}}\,\bigg (\frac{1}{v_q}\bigg)^2 {\rm Re} \bigg
\lbrack \frac {\tilde G_2^{(3)}(s)}{\tilde F_0(s)}\bigg \rbrack
\label{a38a}\\
&=&\frac{1}{3}\bigg (\frac{1}{v_q}\bigg)^2(M/10)^2
\sum_{1\ne k\ne j\ne 1}
\langle \bmF_1(1,j).\bmF_1(1,k)\rangle
\label{a38b}
\end{eqnarray}
\end{mathletters}
involves  the forces  on $'1'$  by two  different medium  particles.  The
expectation value in (\ref{a38b}) requires a  diagonal 3-particle density
matrix, and consistency  requires it to be  the non-diagonal (\ref{a25}),
used in  the calculation of $\tilde\Omega$  in the limit $s=0$. Here too
the 3-body part is negligible.

The  remaining  2-body  parts  may be  compared  with  previous  results
which have been
calculated in  different ways.  Our  result ({\ref{a36}) lies  in between
(0.69,   0.86)   $\AA^{-4}$,   communicated   by   Polls   from   various
approximations  to the  pair-distribution functions  $g(r)$ \cite{polls}.
Another  stochastic  calculation  by  Glyde  and  Boninsegni,
reported  in Ref.  \onlinecite{az1}, leads  to a  result about  35$\%$ in
excess of the above.  Previous  experience has taught that averages, like
the  ones in  (\ref{a31a}) and  (\ref{a31b}) are  quite sensitive  to the
chosen,  pair  distribution.  A  spread  of  10-15$\%$ may  certainly  be
expected, but  presumably not  a deviation of  $\approx 35\%$.

Finally, we compare the computed  total
$q^{*2}\bar\beta_4(q)=1.17\AA^{-4}$   with  a   few
results, extracted from cumulant analyses of the data.
For instance in Ref. \onlinecite{ken} a value compatible with 0
is  given, while  an  upper limit  $q^{*2}\bar\beta_4(q)<0.50 A^{-4}$  is
cited in\ Ref. \onlinecite{az2}.

We close this Section by  comparing $\tilde F_2(s)$, Eq. (\ref{a14}), and
other published  expressions for the same  \cite{grs1,gr,bes}.  Those are
also quadratic in  $V$, but contain in addition  to $\rho_2$, derivatives
of $\rho_2$ and $V$. In contra-distinction our result is quadratic in $V$
and free of derivatives. We have shown above that each of its composing
parts is $\propto \lbrack\bmF_1^2 \rbrack$ with different co-factors. The
alternative expressions provide directly one factor $\bmF_1$ and
it is not  at all evident that the other part can  be cast in that
form.  The equations of motion for density matrices ultimately provide
the evidence. The procedure followed in Eqs. ({\a13})-({\a16})
avoids those
steps  and leads  directly to  the desired  result.  This  can  be
checked for the  general response of a particle in  a potential, Eq. (8c)
of Ref. \onlinecite{gr}.

\section{Summary and conclusion.}

We have derived above an exact expression for the contribution of ternary
collisions to the response of  a non-relativistic many-body system, where
the struck  constituent interacts with  two other medium  particles.  Its
numerical  contribution has  for the  first time  been evaluated  for the
response of liquid $^4$He, $T>T_c$ and momentum transfers in excess of 21
$\Ain$.  For  those we know  that the  asymptotic limit and  the dominant
binary collision correction, accurately describe  the response in a broad
region around  the quasi-elastic  peak, but not  necessarily at  the peak
itself (cf. for instance Ref. \onlinecite{rtmp}).

Our main interest was therefore focused on $y\approx 0$ and the region of
the wings, where  the intensity is only  a fraction of that  in the peak.
Compromising only on  the assumed 3-body density matrix,  we computed the
relative size  of small  TC FSI  effects and  found those  to discernibly
improve the agreement with the data.

The   above   calculation   completes   a  program   to   calculate   the
medium-to-large $q$ response  of liquid $^4$He.  A  number of conclusions
are in  order.  Using  exclusively the well-known  atom-atom interaction,
basic   ground-state  properties   as   are   the  single-atom   momentum
distribution,   the    pair-correlation   function    and   non-diagonal,
two-particle  density  distribution  have   been  determined  with  great
accuracy.

The above  quantities are  then basic  input for  the calculation  of the
linear response of the system.  Only  weak assumptions have been used for
the required two-  and three-particle density matrices,  diagonal in all,
except one coordinate.   Excellent agreement has been  obtained with data
for a theory with demonstrated convergence.

Indeed, given the non-negligible scatter  in the data and observing
that one deals with atomic dynamics and  not with QED, we feel that there
is at present  no incentive to study even finer  theoretical details than
discussed up to now.

Our final remark regards the response of liquid $^4$He when compared with
the  responses of  other systems,  composed of  atoms, molecules,  atomic
nuclei or sub-hadronic matter.  We do  not know of a system where the
approach to the asymptotic limit has been measured and  studied with  an
accuracy, possible for $^4$He.

\section{Acknowledgments.}

ASR thanks A. Polls and H. Glyde for informative discussions.

\newpage

{Figure Captions}

Figs. 1a,b.  The approach to the asymptotic limit $F_0(y)$ (diamonds) for
small $y$  of the calculated even part of the response  without and with
the (even) TC contributions.

Figs. 2a,b
The fractional effect $\alpha(q,y)$, Eq. (\ref{a28}) of TC contributions
in the wings $2.5\lesssim |y|\,({\rm in \,\Ain}) \lesssim 3.3$.

Fig. 3a. Calculated response and data for
$q=21 \Ain, \,T=2.5$ K, including
the effect of instrumental resolution. Dashed and drawn curves are
without, respectively including TC contributions.

Fig. 3b. Same as Fig. 3a for $q=25 \Ain$.

Fig. 3c. Same as Fig. 3a for $q=29 \Ain$.


\begin{references}
\bibitem{foot}

The GRS theory is not a perturbation series in $V$ with
$F_n\propto V^n$, which only holds for $n=1,2$.
\bibitem{foot1}
Eq. (\ref{a29b}) is the correct expression for
$\bar\beta_4$. There are a number of publications
(cf. \cite{glyde,glyde2}) where $\bmF_1(i,j)$ appears replaced by its
component $\hat\bmq.\bmF_1(1,j)$ in the
$\hat\bmq$ direction. For isotropic media
$\langle [\hat\bmq.\bmF_1(1,j)]^2\rangle
=\frac{1}{3}\langle [\bmF_1(1,j)]^2\rangle$.
\end{references}
\end{document}